\documentclass[conference]{IEEEtran}
\IEEEoverridecommandlockouts

\usepackage{cite}
\usepackage{amsmath}
\usepackage{mathtools}
\usepackage{graphicx}
\usepackage[english]{babel}
\usepackage[unicode=true]{hyperref}

\usepackage{flushend}

\usepackage{enumitem}

\usepackage{booktabs}

\usepackage{adjustbox}
\usepackage[table]{xcolor}
\colorlet{tablerowcolor}{gray!20} 
\newcommand{\rowcol}{\rowcolor{tablerowcolor}} %

\begin{document}

\title{
Low-Complexity Own Voice Reconstruction\\for Hearables with an In-Ear Microphone 
\thanks{
The Oldenburg Branch for Hearing, Speech and Audio Technology HSA is funded in the program \frqq Vorab\flqq by the Lower Saxony Ministry of Science and Culture (MWK) and the Volkswagen Foundation for its further development. 
This work was partly funded by the German Ministry of Science and Education BMBF FK 16SV8811 and the Deutsche Forschungsgemeinschaft (DFG, German Research Foundation) - Project\,ID\,352015383 - SFB 1330 C1.}
}

\author{\IEEEauthorblockN{
Mattes Ohlenbusch\IEEEauthorrefmark{1},
Christian Rollwage\IEEEauthorrefmark{1},
Simon Doclo\IEEEauthorrefmark{1}\IEEEauthorrefmark{2}
}
\IEEEauthorblockA{
\small
\IEEEauthorrefmark{1} Fraunhofer Institute for Digital Media Technology IDMT, Oldenburg Branch for Hearing, Speech and Audio Technology HSA, Germany \\
\IEEEauthorrefmark{2} Dept. of Medical Physics and Acoustics and Cluster of Excellence Hearing4all, Carl von Ossietzky Universität Oldenburg, Germany
}
}

\maketitle

\begin{abstract}
Hearable devices, equipped with one or more microphones, are commonly used for speech communication. 
Here, we consider the scenario where a hearable is used to capture the user’s own voice in a noisy environment. 
In this scenario, own voice reconstruction (OVR) is essential for enhancing the quality and intelligibility of the recorded noisy own voice signals. 
In previous work, we developed a deep learning-based OVR system, aiming to reduce the amount of device-specific recordings for training by using data augmentation with phoneme-dependent models of own voice transfer characteristics. 
Given the limited computational resources available on hearables, in this paper we propose low-complexity variants of an OVR system based on the FT-JNF architecture and investigate the required amount of device-specific recordings for effective data augmentation and fine-tuning. 
Simulation results show that the proposed OVR system considerably improves speech quality, even under constraints of low complexity and a limited amount of device-specific recordings.
\end{abstract} 


\section{Introduction}
Speech communication is often impaired in noisy environments. 
In-the-ear hearable devices, i.e., smart earpieces with a loudspeaker and one or more microphones, can be used to improve communication in such environments.
Here, we consider the scenario where a hearable with an outer and an in-ear microphone aims to capture the user's own voice, e.g., to be transmitted via a wireless link to another hearable or a mobile phone. 
The outer microphone captures environmental noise along with recording the own voice.
While the in-ear microphone benefits from the attenuation of environmental noise due to ear canal occlusion, the recorded own voice suffers from low-frequency amplification (below ca.~1\,kHz), band-limitation (above ca.~2\,kHz), and body-produced noise~\cite{bouserhal_-ear_2019}.
The goal of own voice reconstruction (OVR) is to estimate clean broadband own voice signals from the outer and/or in-ear microphone signals. 

%
Several OVR approaches have been proposed which extend the bandwidth of the in-ear microphone signal~\cite{ohlenbusch_training_2022, hauret_configurable_2023, edraki_speaker_2024, sui_tramba_2024, li_restoration_2024, li_two-stage_2024}\footnote{
Although some of these approaches have been proposed and validated for body-conduction microphones, they can also be applied to in-ear microphones.
}.
However, it has been shown in~\cite{yu_time-domain_2020, wang_multi-modal_2022, wang_fusing_2022, ohlenbusch_multi-microphone_2024, ohlenbusch_speech-dependent_2024} that speech quality can be further improved by using outer microphones in addition to in-ear microphones.
Although previously proposed deep learning-based OVR systems 
often have high computational complexity and millions of parameters, it is crucial that OVR systems for hearables have low complexity and few parameters to meet hardware requirements.
In addition, training an OVR system typically requires a large amount of device-specific own voice signals. 
Whereas some OVR approaches only use device-specific own voice recordings directly as training data, e.g.,~\cite{wang_fusing_2022, jhauret-et-al-2024-vibravox}, 
other approaches perform training with augmented own voice data generated from a small amount of device-specific recordings and then perform fine-tuning with the recorded own voice signals~\cite{ohlenbusch_training_2022, edraki_speaker_2024, sui_tramba_2024, ohlenbusch_speech-dependent_2024}.
For single-channel speech enhancement systems, the amount of required training data tends to decrease as complexity decreases~\cite{zhang_beyond_2024}.
However, it is unclear if this relationship also applies to training low-complexity OVR systems with augmented own voice data and fine-tuning using only few device-specific own voice recordings.

%
In this paper, we propose a low-complexity OVR system based on the frequency and time joint non-linear filter (FT-JNF) architecture~\cite{tesch_insights_2023}.
We train the OVR system using a phoneme-dependent own voice data augmentation method proposed in~\cite{ohlenbusch_speech-dependent_2024}. 
For several variants of the proposed OVR system, differing in size and computational complexity, we compare the OVR performance with baseline systems. 
In addition, we investigate the influence of the amount of device-specific own voice recordings used for data augmentation and fine-tuning on the OVR performance.
Experimental results show that the proposed system outperforms baseline systems at a comparable complexity, even when only a small amount of device specific recordings is available.

\section{Signal model}
\begin{figure}
    \centering
    \includegraphics[width=\columnwidth]{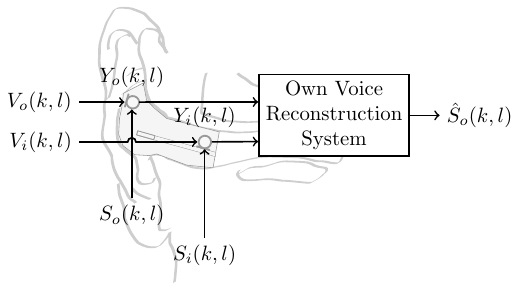} 
    \caption{Block diagram of own voice reconstruction using an outer and an in-ear microphone of a hearable.}
    \label{fig:signal_model_diagram}
\end{figure}
We consider a hearable device equipped with an outer microphone and an in-ear microphone, as depicted in Fig.~\ref{fig:signal_model_diagram}.
The signals are denoted by subscripts $o$ for the outer microphone and $i$ for the in-ear microphone. 
In the short-time Fourier transform (STFT) domain, $S_o(k,l)$ and $S_i(k,l)$ denote the own voice signals of the user at both microphones, where $k$ and $l$ denote the frequency index and the frame index.
The outer and in-ear microphone signals are given by 
\begin{align}
    Y_o(k,l) = & S_o(k,l) + V_o(k,l),\\
    Y_i(k,l) = & S_i(k,l) + V_i(k,l),
\end{align}
where we assume that the noise components $V_o(k,l)$ and $V_i(k,l)$ predominantly consist of environmental noise.
Self-noise is recorded at both microphones, with additional body-produced noise also being recorded at the in-ear microphone.

\section{Own voice reconstruction system}
\label{sec:ovr_system}
\begin{figure}
\centering
    \includegraphics{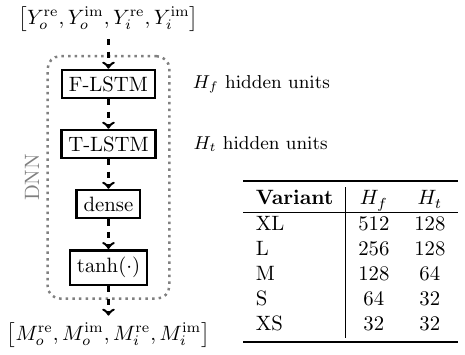}
    \caption{Architecture of the OVR system, estimating complex-valued masks for the outer and in-ear microphones. The number of hidden units, $H_f$ and $H_t$, is different for each proposed variant.}
    \label{fig:dnn_architecture}
\end{figure}

The goal of own voice reconstruction is to estimate the own voice signal $S_o(k,l)$ from the outer and in-ear microphone signals.
In~\cite{ohlenbusch_multi-microphone_2024, ohlenbusch_speech-dependent_2024} an OVR system based on the FT-JNF architecture~\cite{tesch_insights_2023} has been proposed, see Fig.~\ref{fig:dnn_architecture}.
This system takes the complex-valued outer and in-ear microphone STFT coefficients as input, split into real and imaginary parts $Y_o^\text{re}(k,l)$ and $Y_o^\text{im}(k,l)$ for the outer microphone and $Y_i^\text{re}(k,l)$ and $ Y_i^\text{im}(k,l)$ for the in-ear microphone.
The input is processed by a frequency-direction LSTM (F-LSTM) with $H_f$ hidden units, followed by a time-direction LSTM (T-LSTM) with $H_t$ hidden units.
The output of the T-LSTM layer is then processed by a dense layer and a $\tanh$ activation function to obtain the real and imaginary parts of the complex-valued STFT masks $M_o(k,l)$ and $M_i(k,l)$ for the outer and in-ear microphones.
The $\tanh$ activation function constrains the real and imaginary parts of the masks to the range $[-1,~1]$. 
To compute the own voice estimate $\hat{S}_o(k,l)$, noisy STFT coefficients of both microphones are multiplied with the corresponding masks and added, i.e.,
\begin{equation}
    \label{eq:ftjnf_masking}
    \hat{S}_o(k,l) = \sum_{ \mathclap{m\in\{o, i\}} }  M_m(k,l) \cdot Y_m(k,l).
\end{equation}
In this paper we consider several variants of this OVR system, which differ in size and computational complexity.
We vary the size by changing the number of hidden units $H_f$ and $H_t$, see numbers in Fig.~\ref{fig:dnn_architecture} for the extra-large (XL), large (L), medium (M), small (S) and extra-small (XS) variants. 
In this paper, we do not consider further complexity reduction, e.g., by quantization or pruning~\cite{stamenovic_weight_2021}.

\section{Phoneme-dependent own voice augmentation} 
\label{sec:ownvoice_augmentation}
Training an OVR system using both an outer and an in-ear microphone requires a device-specific dataset of own voice signals.
Recording a dataset sufficiently large for direct training requires considerable effort.
However, it is more feasible to record only a small dataset with a limited number of talkers or utterances per talker.
In~\cite{ohlenbusch_modeling_2023}, a method was proposed to simulate in-ear own voice signals from outer microphone signals (for talkers and utterances not included in the small recorded dataset).
From the small recorded dataset, phoneme-specific relative transfer functions (RTFs) between the outer and the in-ear microphone are estimated for each recorded talker, over all time frames in which a specific phoneme $p$ occurs.
The estimated phoneme-specific RTFs are denoted by $\hat{H}_{p}(k)$. 
For RTF estimation, it is assumed here that there is no environmental noise in the small recorded dataset, and that the sensor noise is negligible compared to the own voice in both microphone signals.

For simulation, an outer microphone signal $S_o(k,l)$ of a random different talker is phoneme-annotated to obtain the phoneme annotation sequence $p_o(l)$.
The simulated in-ear signal $\hat{S}_i(k,l)$ is then obtained as
\begin{equation}
    \hat{S}_i(k,l) =  \hat{H}_{p_{o}(l)}(k) \cdot S_o(k,l). 
    \label{eq:augmentation_speechdep}
\end{equation} 
Additionally, to avoid artifacts during phoneme transitions, temporal smoothing of $\hat{H}_{p_{o}(l)}(k)$ is carried out (see~\cite{ohlenbusch_modeling_2023} for details).   
Instead of assuming outer microphone recordings are available, it was proposed in~\cite{ohlenbusch_speech-dependent_2024} to use clean speech signals from standard datasets instead.
Since standard datasets are readily available, this method allows for the simulation of a large amount of simulated in-ear own voice signals. 
An OVR system can then be trained with augmented own voice signals, consisting of clean speech signals used as the outer microphone own voice signal and the corresponding simulated in-ear own voice signal.
In~\cite{ohlenbusch_speech-dependent_2024} it was shown that when a small dataset of device-specific recordings is available, this data augmentation method can improve OVR performance compared to only using the device-specific recordings directly as training data. 
After training an OVR system with augmented own voice signals, the recorded own voice signals can be used to fine-tune the system, further improving performance.

\section{Experimental setup} 
To evaluate the proposed FT-JNF variants and several baseline systems (see Section~\ref{sec:baselines}) for own voice reconstruction, we conduct an experimental evaluation.
In this section, we describe the experimental setup for the evaluation.

\subsection{Datasets}
\label{sec:data}
The evaluation uses own voice recordings made with the Hearpiece hearable prototype~\cite{denk_one-size-fits-all_2019}.
A dataset of German own voice signals of 18 talkers with 306 utterances each is split into training, validation, and test sets with 12, 2, and 4 talkers, respectively.
All OVR systems are first trained on augmented own voice signals and then fine-tuned on recorded own voice signals.
The augmented own voice signals are obtained by augmenting 10\% of the German portion of the CommonVoice dataset~\cite{ardila_common_2020} (v11.0), corresponding to 115.7 hours, as described in Section~\ref{sec:ownvoice_augmentation}.
The full augmented training and fine-tuning of the proposed variants and the baseline systems uses own voice recordings from 12 talkers with 306 utterances each. Reduced amounts are considered in Section~\ref{sec:results_resource}. 
It should be noted that independent of the amount of used device-specific recordings, all systems were trained with the same amount of augmented data (115.7 hours).
The noise signals used for training and testing are a spatialized version of the fifth DNS challenge~\cite{dubey_icassp_2024}, obtained following the procedure in~\cite{ohlenbusch_multi-microphone_2024} using individually matched, measured transfer functions for the same users as in the recorded own voice dataset\footnote{German own voice dataset [online]: \url{https://doi.org/10.5281/zenodo.10844599}, individual transfer function measurements [online]: \url{https://doi.org/10.5281/zenodo.11196867}}. Measurements from 8 horizontal directions in 45°-steps with 1.5\,m distance are used to compute either point source signals (single direction) or pseudo-diffuse noise signals (8 directions). 

\subsection{Training details} 
\label{sec:training_details}
The experiments are conducted at a sampling rate of 16\,kHz, using an STFT framework with a frame length of 32\,ms and a frame shift of 16\,ms, where a square-root Hann window is used both in analysis and synthesis. 
Own voice and noise signals are mixed at a random signal-to-noise ratio (SNR) between -10 and 25\,dB, defined at the outer microphone. 
Training is carried out with four examples per batch and an example length of 3\,s, using the combined $L_1$ loss between the target clean own voice signal at the outer microphone and the estimated own voice signal in the time domain and the STFT domain (after re-analysis)~\cite{wang_stft-domain_2023}.
The ADAM optimizer~\cite{kingma_adam_2015} is used with an initial learning rate of $10^{-4}$, which is halved after three epochs without improvement of the validation loss, and training is stopped after six epochs without improvement. 
The initial learning rate for fine-tuning is $10^{-5}$.

\subsection{Evaluation metrics} 
\label{sec:eval_metrics}
OVR performance is evaluated using wideband PESQ~\cite{international_telecommunications_union_itu_itu-t_2001}, extended short-time objective intelligibility (ESTOI)~\cite{jensen_algorithm_2016}, and log-spectral distance (LSD)~\cite{gray_distance_1976}. 
For all three metrics, the clean own voice signal at the outer microphone is chosen as the reference signal.
Higher PESQ and ESTOI values are better, while lower LSD values are better. 
During testing, OVR performance is evaluated at SNRs of -10, -5, 0, 5, and 10\,dB.
The results are averaged over the test set and over SNR.
System complexity is reported in terms of number of parameters, number of multiply-accumulate operations per second (MACs/s), and real-time factor (RF).
MACs are computed using the \texttt{thop} Python package.
The RF is computed on an Intel Core i7-10850H CPU (2.7\,GHz).

\subsection{Baseline systems}
\label{sec:baselines}
The baseline systems include three systems that only use the in-ear microphone (IM) signal, and one system that uses both the outer and in-ear microphone signals.
All baseline systems were retrained using the same setup as described in Section~\ref{sec:training_details} for the proposed FT-JNF variants:
\begin{itemize}
    \item UNet~\cite{ohlenbusch_training_2022,wang_towards_2021}: 
    Time-domain system performing reconstruction of the in-ear own voice signal.
    \item Extreme Bandwidth Extension Network (EBEN)~\cite{hauret_configurable_2023}: 
    Time-domain system, originally proposed for bandwidth extension of body-conducted speech. 
    The generator was retrained by replacing the generative adversarial network-based training with the loss function from~\cite{wang_stft-domain_2023}, as used for all other systems.
    \item FT-JNF~XL (IM): 
    The proposed FT-JNF~XL using only the in-ear microphone signal. 
    Due to the activation function and only estimating masks (see Section~\ref{sec:ovr_system}), this system is unable to perform bandwidth extension. 
    \item Group Communication Binaural Filter and Sum Network (GCBFSNet)~\cite{westhausen_binaural_2023}: Unilateral version of the low-complexity GCBFSNet (8 groups, 32 hidden units, with post-filter, 2\,ms frames, 1\,ms frame shift), retrained for OVR using both the outer and in-ear microphone signals.
\end{itemize}

\section{Results}
In this section, the results of the experimental evaluation are presented.
In Section~\ref{sec:results_compare_baselines}, the proposed FT-JNF variants are compared to the baseline systems.
In Section~\ref{sec:results_resource}, the influence of the amount of device-specific own voice recordings for data augmentation and fine-tuning is investigated.
Audio examples from the evaluation are available online\footnote{Audio examples [online]: \url{https://m-ohlenbusch.github.io/low_complexity_ovr_examples/}}.

\subsection{Comparison to baseline systems}
\label{sec:results_compare_baselines}

\begin{table} 
    \caption{Performance, size and complexity of the baseline systems and the proposed FT-JNF variants (XL, L, M, S, XS). 
    'M' indicates millions, 'G' indicates billions. 
    Rows with a gray background indicate systems using only the in-ear microphone.
    }
    \label{tab:results_bigtable}
    \centering
  \begin{adjustbox}{max width=\columnwidth}
   \begin{tabular}{l|rrr|rrr}
    \toprule
    \multicolumn{1}{l|}{} & \multicolumn{3}{c|}{Intrusive metrics} & \multicolumn{3}{c}{Size and complexity}  \\
    \hline
    \textbf{System} & \textbf{PESQ} & \textbf{ESTOI} & \textbf{LSD} & \textbf{Param.} & \textbf{MACs/s} & \textbf{RF} \\ 
    \hline
    Unprocessed                                          & 1.25 & 0.51 & 2.46 & -        &        - &     -  \\ 
    \hline
    \rowcol UNet (IM)~\cite{ohlenbusch_training_2022}    & 1.85 & 0.65 & 1.30 & 10.278\,M &  6.03\,G & 0.157 \\
    \rowcol EBEN (IM)~\cite{hauret_configurable_2023}    & 1.51 & 0.57 & 1.64 &  1.946\,M &  1.02\,G & 0.034 \\ 
    \rowcol FT-JNF XL (IM)                               & 1.47 & 0.61 & 1.73 &  1.390\,M & 22.38\,G & 0.387 \\ 
    GCBFSNet~\cite{westhausen_binaural_2023}             & 1.93 & 0.68 & 1.36 &  0.100\,M &  0.31\,G & 0.303 \\
    \hline
    FT-JNF XL & 2.58 & 0.78 & 1.08 & 1.390\,M & 22.45\,G & 0.392 \\
    FT-JNF L  & 2.50 & 0.77 & 1.10 & 0.466\,M &  7.55\,G & 0.173 \\
    FT-JNF M  & 2.22 & 0.72 & 1.27 & 0.118\,M &  1.93\,G & 0.071 \\
    FT-JNF S  & 2.18 & 0.72 & 1.28 & 0.031\,M &  0.50\,G & 0.029 \\
    FT-JNF XS & 1.95 & 0.68 & 1.40 & 0.013\,M &  0.23\,G & 0.011 \\ 
    \bottomrule
    \end{tabular}
    \end{adjustbox}
\end{table}
Table~\ref{tab:results_bigtable} compares the performance, size and complexity of the proposed FT-JNF variants and the baseline systems.
First, it can be observed that all OVR variants achieve considerable improvements in all metrics compared to the unprocessed (noisy outer microphone) signals.
Not surprisingly, systems using both the outer microphone and the in-ear microphone (GCBFSNet and FT-JNF variants) generally outperform systems using only the in-ear microphone (UNet, EBEN, FT-JNF~XL (IM)).
Among the systems using only the in-ear microphone, UNet achieves the best scores but also has the most parameters. 
While EBEN and FT-JNF~XL (IM) have a similar amount of parameters and performance, FT-JNF~XL (IM) has a much higher complexity (MACs/s and RF).
Among the systems using both the outer and the in-ear microphone, GCBFSNet has a slightly lower RF than the FT-JNF~XL variant, but higher than the L, M, S, XS variants.
Although GCBFSNet has fewer MACs/s than FT-JNF~S, FT-JNF~S has about three times fewer parameters and achieves better scores in all metrics. FT-JNF~XS performs comparable to GCBFSNet with fewer MACs/s and at a much lower RF.

FT-JNF~XL performs much better than IM-FT-JNF~XL, while the complexity of FT-JNF~XL is only marginally higher.
This indicates a substantial performance gain from using the outer microphone. 
Due to performing masking in a constrained value range, FT-JNF~XL (IM) is unable to reconstruct speech in high frequency regions, whereas FT-JNF~XL can use high frequency content from the outer microphone.

While FT-JNF~XL consists of 1.39 million parameters, it requires a high number of computations due to the F-LSTM iterating over all frequencies for each time frame.
When the model complexity is decreased to L, the MACs/s and RF also decrease, while the performance only slightly decreases. 
Even though the performance of the smaller variants (M, S and XS) is lower compared to FT-JNF~XL and L, their performance is still better than the baseline systems. 
It should be noted that FT-JNF~S and XS require approximately 44 and 97 times fewer MACs/s than FT-JNF~XL, respectively.

\subsection{Influence of amount of device-specific recordings} 
\label{sec:results_resource}
\begin{figure} 
    \centering
    \includegraphics
    {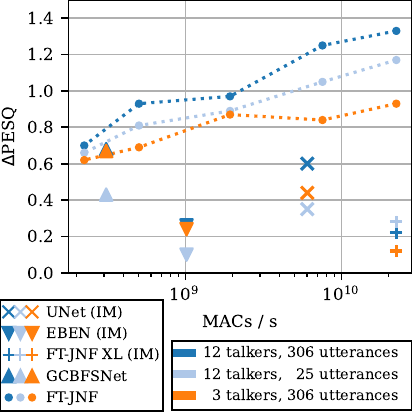} 
    \caption{PESQ improvement of the baseline systems and the proposed FT-JNF variants for different amounts of device-specific recordings (talkers, utterances). 
    Different systems are distinguished by different symbols, while different amounts of recordings are represented by different colors. 
    }
    \label{fig:pesq_results} 
\end{figure}
To investigate the relationship between system complexity and amount of device-specific recordings, the baseline systems and the proposed FT-JNF variants were retrained using different amounts of device-specific recordings for augmented training and fine-tuning.
We investigated both the influence of reducing the number of talkers from 12 to 3 (with 306 utterances) and reducing the number of utterances from 306 to 25 (for 12 talkers).
Fig.~\ref{fig:pesq_results} shows the results in terms of PESQ improvement ($\Delta$PESQ) compared to the unprocessed noisy outer microphone signals. 
When the number of talkers is reduced, the performance of baselines with low complexity (GCBFSNet, EBEN) only slightly decreases while for UNet and FT-JNF~XL (IM) there is a larger decrease.  
For the proposed variants, a large performance decrease from a reduced number of talkers is observed for the XL and L variants, while the performance only slightly decreases for the M, S, and XS variants.
When the number of recorded utterances per talker is reduced, for the baselines the performance decrease is larger than when reducing the number of talkers, but it is smaller for the proposed variants. 
The results indicate that low-complexity OVR systems require fewer device-specific recordings for augmented training and fine-tuning than systems with higher computational complexity.

\section{Conclusion}
In this paper, we proposed variants of the FT-JNF architecture with low computational complexity for OVR.
We investigated the influence of the amount of device-specific recordings used for data augmentation and fine-tuning on the OVR performance.
Experimental results demonstrate that the proposed variants outperform baseline systems at a comparable complexity. 
Even under constraints of low complexity and a limited amount of device-specific recordings available for training, considerable quality improvements can be achieved by the proposed system.

\section*{Acknowledgement}
The authors would like to thank Nils~L.~Westhausen for providing the code for the original GCBFSNet architecture.

\clearpage 
\bibliographystyle{IEEEbib}
\bibliography{manualbib}

\end{document}